\documentclass[aps,pra,preprint,groupedaddress,showpacs]{revtex4}
\usepackage{amsmath,amssymb,amsthm,amsfonts}
\usepackage{graphicx}
\usepackage{bm}
\begin{document}
\title{ Low-energy three-body dynamics in binary quantum gases 
}

\author{O.~I.~Kartavtsev}
\author{A.~V.~Malykh}
\affiliation{ Joint Institute for Nuclear Research, Dubna, 141980, Russia }


\begin{abstract}

The universal three-body dynamics in ultra-cold binary Fermi and Fermi-Bose 
mixtures is studied. 
Two identical fermions of the mass $m$ and a particle of the mass $m_1$ with 
the zero-range two-body interaction in the states of the total angular 
momentum $L = 1$ are considered. 
Using the boundary condition model for the $s$-wave interaction of different 
particles, both eigenvalue and scattering problems are treated by solving 
hyper-radial equations, whose terms are derived analytically. 
The dependencies of the three-body binding energies on the mass ratio $m/m_1$ 
for the positive two-body scattering length are calculated; it is shown that 
the ground and excited states arise at $m/m_1 \ge \lambda_1 \approx 8.17260$ 
and $m/m_1 \ge \lambda_2 \approx 12.91743$, respectively. 
For $m/m_1 \alt \lambda_1$ and $m/m_1 \alt \lambda_2$, the relevant bound 
states turn to narrow resonances, whose positions and widths are calculated. 
The 2 + 1 elastic scattering and the three-body recombination near 
the three-body threshold are studied and it is shown that a two-hump structure 
in the mass-ratio dependencies of the cross sections is connected with arising 
of the bound states. 

\end{abstract}


\maketitle

\section{Introduction} 
\label{Introduction}

In the last years, investigations of multi-component ultra-cold quantum gases 
have attracted much interest. 
Properties of binary Fermi-Bose~\cite{Ospelkaus06,Karpiuk05} and 
Fermi~\cite{Shin06,Chevy06,Iskin06} mixtures and of impurities embedded in 
a quantum gas~\cite{Cucchietti06,Kalas06}) are under experimental and 
theoretical study. 
Different aspects of the few-body dynamics of two-species compounds are of 
interest both from the general point of view and for many-body applications. 
For example, there are an infinite number of three-body bound states of 
two-component fermions (Efimov effect) if their mass ratio exceeds 
the critical value~\cite{Efimov73}. 
Recently, an infinite number of $1^+$ bound states has been 
predicted~\cite{Macek06} for three identical fermions with the resonant 
$p$-wave interaction. 
More detailed study of the energy spectrum of three two-component particles 
is of interest to shed light on the role of trimer molecules in the many-body 
dynamics and provides an insight into the few-body processes. 
Concerning the low-energy scattering, one of the interesting features is 
a two-hump structure in the isotopic dependence of the three-body 
recombination rate of two-component fermions~\cite{Petrov03,Petrov05a}. 
Note that the low-energy three-body recombination rate of two-component 
fermions scales as the first power of the collision energy and the sixth power 
of the two-body scattering length~\cite{Petrov03,Suno03}. 

The aim of the present paper is to study the three-body energy spectrum and 
the low-energy (2 + 1)-scattering for two identical fermions of mass $m$ and 
the third different particle of mass $m_1$. 
Here one considers the unit total angular momentum $L = 1$ and the $s$-wave 
interaction between different particles, which is most important for 
description of the low-energy processes. 
Note that the $s$-wave interaction takes place only in binary mixtures, 
whereas only the $p$-wave interaction is possible in a one-component Fermi 
gas. 
The description of the three-body properties turns out to be universal and 
depending on a single parameter $m/m_1$ in the limit of the zero interaction 
range. 
For the interaction given within the framework of the boundary condition model 
(BCM), solution of hyper-radial equations (HREs) provides an efficient 
approach to treat both the eigenvalue and the scattering 
problem~\cite{Kartavtsev99,Nielsen99,Kartavtsev02}. 
An important advantage of the BCM is that all the terms of HREs are derived in 
the analytical form; the method of derivation and the analytical expressions 
are similar to those obtained for three identical bosons in 3 and 2 
dimensions~\cite{Kartavtsev99,Kartavtsev06}. 
The calculations reveal that two three-body bound states arise while the mass 
ratio $m/m_1$ increases from zero to the critical $\lambda_c$, beyond which 
the number of bound states becomes infinite~\cite{Efimov73}. 
A two-hump structure is found for the mass-ratio dependencies of the elastic 
and inelastic (2 + 1)-scattering cross sections near the three-body threshold. 
The structure of the isotopic dependencies is qualitatively related to arising 
of the bound states as a common origin is an increase of the potential-well 
depth in the 2 + 1 channel. 

\section{Outline of the approach}
\label{approach}

In the universal low-energy limit, the short-range two-body interaction is 
described by a single parameter, a natural choice for which is the two-body 
scattering length $a$.  
For the vanishing range of interaction, the two-body interaction is defined 
within the framework of the BCM by imposing the boundary condition at the zero 
inter-particle distance $r$ 
\begin{eqnarray}
\label{bound1}
\lim_{r \rightarrow 0}\frac{\partial \ln (r\Psi)} {\partial r} = 
- \frac{1}{a}\ .
\end{eqnarray}
The two-body interaction introduced in this way is known in the literature as 
the zero-range potential~\cite{Demkov88}, the Fermi 
pseudo-potential~\cite{Wodkiewicz91}, the Fermi-Huang 
pseudo-potential~\cite{Idziaszek06,Kanjilal06}, and an equivalent approach is 
used in the momentum-space representation~\cite{Braaten03}. 

For definiteness, one supposes that particle 1 is of mass $m_1$ and particles 
2 and 3 are two identical fermions of mass $m$. 
The wave function $\Psi$ satisfies the equation 
\begin{equation}
\label{shred} 
\left[\Delta_{{\mathbf x}} + \Delta_{{\mathbf y}} + E\right]\Psi = 0 \ ,
\end{equation}
where the scaled Jacobi variables ${\mathbf x} = 
\displaystyle\sqrt{2\mu}\left({\mathbf r}_2 - {\mathbf r}_1\right)$ and 
${\mathbf y} = \displaystyle\sqrt{2\tilde\mu}\left({\mathbf r}_3 - 
\frac{m_1{\mathbf r}_1 + m{\mathbf r}_2}{m_1 + m}\right)$ are defined via 
the position vectors ${\mathbf r}_i$ and the reduced masses 
$\mu = \displaystyle\frac{mm_1}{m + m_1}$ and $\tilde{\mu} = 
\displaystyle\frac{m(m + m_1)}{m_1 + 2m}$. 
The total interaction is expressed by imposing two boundary conditions of 
the form~(\ref{bound1}) at zero distances between the different particles in 
two pairs $1 - 2$ and $1 - 3$. 
One demands that the wave function should be antisymmetric under permutation 
of identical fermions 2 and 3; under this condition only a single boundary 
condition of the form~(\ref{bound1}) should be imposed at the zero distance 
between particles $1$ and $2$, $x \to 0$. 
The unit total angular momentum $L = 1$ is considered, which is most important 
for the low-energy processes~\cite{Petrov03,Suno03}. 
The units $\hbar = 2\mu = |a| = 1$ will be used hereafter; thus, any 
three-body property depends only on the single remaining parameter $m/m_1$. 

The three-body bound and resonance states and the low-energy scattering are 
conveniently treated by solving a system of HREs~\cite{Macek68}. 
The eigenfunctions $\Phi_n(\rho, \Omega)$ are defined as regular solutions on 
the hypersphere at the fixed hyper-radius $\rho$, 
\begin{eqnarray}
\label{eqonhypershere}
&& \left[\frac{1}{\sin^2 2\alpha}\left( \sin^2 2\alpha
\frac{\partial}{\partial \alpha} \right) + \frac{1}{\sin^2\alpha}
\Delta_{\hat{\mathbf x}} + \frac{1}{\cos^2\alpha}\Delta_{\hat{\mathbf y}} + 
\gamma^2_n(\rho) - 4\right]\Phi_n(\rho, \Omega) = 0 \ , \\
\label{bch}
&& \lim_{\alpha\rightarrow 0} 
\left[ \frac{\partial\ln\left(\alpha \Phi_n \right)}{\partial\alpha} \pm \rho 
\right] = 0 \ , 
\end{eqnarray} 
where $\Omega$ is a brief notation for the hyper-angles $\alpha$, 
$\hat{\mathbf x}$, and $\hat{\mathbf y}$. 
The hyper-spherical variables are defined by the relations 
$x = \rho\sin \alpha$, $y = \rho\cos \alpha$, 
$\hat{\mathbf x} = {\mathbf x}/x$, and $\hat{\mathbf y} = {\mathbf y}/y$. 
The $\pm$ sign in~(\ref{bch}), which corresponds to the positive and negative 
scattering length $a$, hereafter will be incorporated into the parameter 
$\rho$. Thus, the eigenvalue problem will be considered for an arbitrary 
$-\infty < \rho < \infty $. 
For each value of $\rho$, the solution of (\ref{eqonhypershere}) 
and~(\ref{bch}) determines a set of discrete eigenvalues $\gamma_n^2(\rho)$, 
which are enumerated in ascending order by an index $n = 1, 2, 3, \dots$, and 
corresponding eigenfunctions $\Phi_n(\rho, \Omega)$. 
The expansion of the total wave function in a set of eigenfunctions 
$\Phi_n(\rho, \Omega)$ normalized by the condition 
$\langle\Phi_n|\Phi_m\rangle = \delta_{nm}$, 
\begin{equation}
\label{Psi} \Psi = \rho^{-5/2} \sum_{n=1}^{\infty}
f_n(\rho)\Phi_n(\rho,\Omega) \ ,
\end{equation}
leads to an infinite set of coupled HREs 
\begin{equation}
\label{system1} 
\left[\frac{d^2}{d \rho^2} - \frac{\gamma_n^2(\rho) - 1/4}{\rho^2} + E \right] 
f_n(\rho) - \sum_{m = 1}^{\infty}\left[P_{mn}(\rho) - Q_{mn}(\rho) 
\frac{d}{d\rho} - \frac{d}{d\rho}Q_{mn}(\rho) \right] f_m(\rho) = 0 \ , 
\end{equation}
where 
\begin{equation}
\label{QPnm0}
Q_{nm}(\rho) = \left\langle\Phi_n \biggm| 
\frac{\partial\Phi_m}{\partial\rho}\right\rangle \ , \quad 
P_{nm}(\rho) = \left\langle\frac{\partial\Phi_n}{\partial\rho} \biggm| 
\frac{\partial\Phi_m}{\partial\rho}\right\rangle \ , 
\end{equation} 
and the notation $\langle\cdot|\cdot\rangle$ stands for integration over 
the invariant volume on the hypersphere $d\Omega = 
\sin^2{2\alpha}\,d\alpha\,d{\hat{\mathbf x}} d{\hat{\mathbf y}}$. 

The eigenfunctions $\Phi_n(\rho, \Omega)$ inherit the antisymmetry of 
the wave function under permutation of the identical fermions 2 and 3. 
The solutions of the eigenvalue problem~(\ref{eqonhypershere}) 
and~(\ref{bch}), which satisfy the permutational symmetry and belong to 
the total angular momentum $L = 1$ and its projection $M$, are expressed as 
\begin{equation} 
\label{faddeev}
\Phi_n(\rho,\Omega) = \left(1 - \widehat{P}\right)
\frac{\varphi_n(\alpha, \rho)}{\sin 2\alpha} Y_{1M}(\hat{\mathbf y}) \ ,
\end{equation}
where $Y_{LM}(\hat{\mathbf y})$ is the spherical function and $\widehat{P}$ 
denotes the permutation of particles 2 and 3. 
The action of $\widehat{P}$ in terms of the Jacobi variables is given by  
\begin{eqnarray}
\label{xy}
\widehat{P}{\mathbf x} = -\sin\omega{\mathbf x} + \cos\omega{\mathbf y} \ , 
\quad 
\widehat{P}{\mathbf y} = -\cos\omega{\mathbf x} - \sin\omega{\mathbf y} \ , 
\end{eqnarray} 
where the angle of the kinematic rotation $\omega $ is expressed via the mass 
ratio as $\cot \omega = \displaystyle(m_1/m)\sqrt{1 + 2 m/m_1}$. 
Given the representation~(\ref{faddeev}), the eigenvalue 
problem~(\ref{eqonhypershere}) and~(\ref{bch}) is reduced to the equation 
\begin{equation}
\label{eqonhyp1}
\left[\frac{\partial^2}{\partial \alpha^2} - \frac{2}{\cos^2\alpha}
 + \gamma^2_n(\rho)\right]\varphi_n(\alpha,\rho) = 0 
\end{equation} 
complemented by the boundary conditions $\varphi(\alpha, \rho) = 0$ at 
$\alpha = \pi/2$ and 
\begin{eqnarray}
\label{bconhyp}
\lim_{\alpha\rightarrow 0} 
\left(\frac{\partial}{\partial\alpha} + \rho \right)\varphi_n(\alpha, \rho) 
+ \frac{2}{\sin 2\omega}\varphi_n(\omega, \rho) = 0 \ , 
\end{eqnarray} 
at the singular point $\alpha = 0$ of the eigenfunctions 
$\Phi_n(\rho,\Omega)$. 
The boundary condition~(\ref{bconhyp}) is deduced from 
Eqs.~(\ref{bch}),~(\ref{faddeev}), and~(\ref{xy}) by observing that 
$\widehat{P}\alpha \to \pi/2 - \omega$ and 
$\widehat{P}Y_{1M}(\hat{\mathbf y}) \to -Y_{1M}(\hat{\mathbf y})$ as 
$\alpha \to 0$. 

The zero-valued at $\alpha = \pi /2$ unnormalized solutions to 
Eq.~(\ref{eqonhyp1}) are straightforwardly written as 
\begin{eqnarray}
\label{varphi}
\varphi_n(\alpha, \rho) = \gamma_n(\rho)\cos\left[\gamma_n(\rho)
\left(\pi/2 - \alpha\right)\right] - \tan\alpha \sin\left[\gamma_n(\rho) 
\left(\pi/2 - \alpha\right)\right] \ . 
\end{eqnarray}
Substituting~(\ref{varphi}) into the boundary condition~(\ref{bconhyp}), 
one eventually finds the transcendental equation, 
\begin{eqnarray}
\label{transeq}
\rho = \frac{1 - \gamma^2}{\gamma}\tan\gamma\frac{\pi}{2} - 
\frac{2}{\sin2\omega} \frac{\cos\gamma\omega}{\cos\gamma\frac{\pi}{2}} + 
\frac{\sin\gamma\omega}{\gamma\sin^2\omega\cos\gamma\frac{\pi}{2}} \ ,
\end{eqnarray}
which determines the infinitely multivalued function $\gamma^2(\rho)$ of 
an arbitrary complex-valued variable $\rho$ at various mass ratios $m/m_1$ 
given by the parameter $\omega$. 
Different branches of this unique function for the real-valued $\rho$ form 
a set of real-valued eigenvalues $\gamma_n^2(\rho)$; thus, the solution of 
the eigenvalue problem is accomplished by means of expressions~(\ref{varphi}) 
and~(\ref{transeq}). 

An advantage of the BCM is that the eigenvalues $\gamma_n^2 (\rho)$ entering 
into HREs are expressed in a simple analytical 
form~\cite{Nielsen99,Kartavtsev02,Kartavtsev06}, which is helpful both for 
qualitative analysis and in the numerical calculations. 
Moreover, the coupling terms $Q_{nm}(\rho) $ and $P_{nm}(\rho) $ can be 
determined in the analytical form via $\gamma_n^2 (\rho)$ and their 
derivatives as was done in~\cite{Kartavtsev99,Kartavtsev06}; the derivation is 
outlined in the Appendix. 

Properties of the eigenvalues $\gamma_n^2(\rho)$ are deduced by analyzing 
Eq.~(\ref{transeq}), in particular, all the $\gamma_n^2(\rho)$ monotonically 
decrease within the intervals $9 > \gamma_1^2(\rho) > - \infty $ and 
$(2n + 1)^2 > \gamma_n^2(\rho) > (2n - 1)^2 $ for $n \ge 2$ as 
the hyper-radius runs the interval $-\infty < \rho < \infty$. 
The effective potentials in the upper channels for $n \ge 2$ contain 
the repulsive term $\gamma_n^2(\rho)/\rho^2$, which means a dominant role of 
the lowest channel for the low-energy solutions. 
Furthermore, the first-channel potential at small $\rho$ is approximately 
determined by $\gamma_1^2(0)$ so that $V_1(\rho) \approx \left[ \gamma_1^2(0) 
- 1/4 \right] /\rho^2$, which entails that a number of the bound states is 
finite for $\gamma_1^2(0) > 0$ and infinite for $\gamma_1^2(0) < 0$. 
As follows from Eq.~(\ref{transeq}), $\gamma_1^2(0)$ decreases with increasing 
$\omega $ and crosses zero at the critical value 
$\omega_c \approx 1.19862376 $, which satisfies the equation 
\begin{eqnarray}
\label{omegacr}
\frac{\pi}{2}\sin^2\omega_c - \tan\omega_c + \omega_c = 0 \ , 
\end{eqnarray}
and corresponds to the critical mass ratio $\lambda_c \approx 13.6069657$.
Thus, one concludes that a number of three-body bound states is either finite 
or infinite if the mass ratio $m/m_1$ is below or above $\lambda_c$. 
An infinite energy spectrum of three fermions for the mass ratio above 
the critical value, $m/m_1 > \lambda_c$, was first discovered in 
Ref.~\cite{Efimov73}. 
Note that the unambiguous description of the three-body properties for 
$m/m_1 > \lambda_c$ requires an additional parameter which determines the wave 
function in the vicinity of the triple-collision point. 

\begin{figure}[hbt]
\includegraphics[width=.7\textwidth]{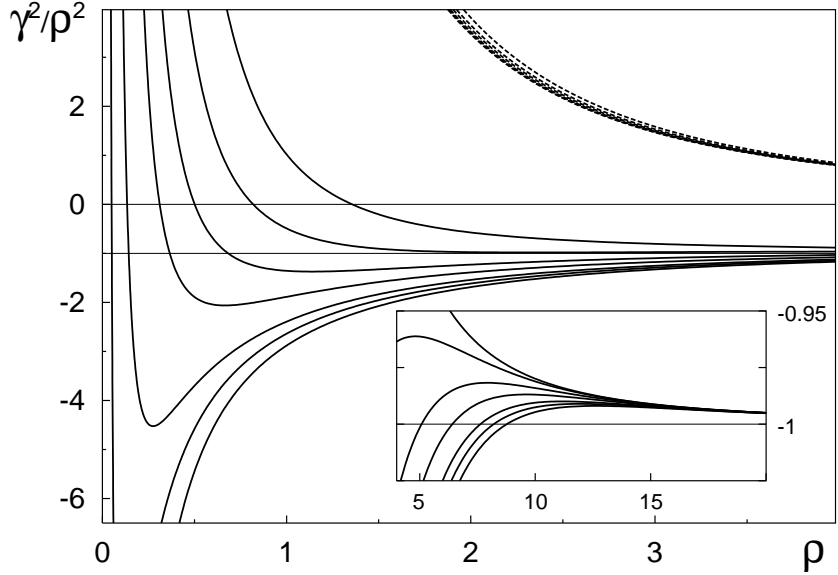} \\ 
{\caption{Diagonal terms in HREs $\gamma^2_1(\rho)/{\rho^2}$ (solid lines) and 
$\gamma^2_2(\rho)/{\rho^2}$ (dashed lines) for a set of mass ratios 
$m/m_1 = 1, 5, 8, 10, 12, 13, 14$ (top to bottom). 
In the inset the first-channel terms are shown on a large scale in the barrier 
region. 
For reference, the two-body threshold at $E = -1$ is plotted. } 
\label{figterm}}
\end{figure}

More detailed analysis is needed to describe the energy spectrum for the mass 
ratio below the critical value, $m/m_1 \le \lambda_c$. 
The description is quite simple for the negative two-body scattering length 
$a < 0$, in which case all the $\gamma_n^2(\rho)$ are non-negative for 
$m/m_1 \le \lambda_c$, which means unboundedness of the three particles. 
Considering the positive two-body scattering length $a > 0$, one finds that 
there are no bound states for the small enough mass ratio roughly below 5 
because the first-channel diagonal term $\gamma_1^2(\rho)/\rho^2 $ exceeds 
the threshold energy $E = -1$ for those $m/m_1 \alt 5$. 
To estimate the number of the bound states which occur as $m/m_1$ increases to 
the critical mass ratio $\lambda_c$, one should consider the small-$\rho$ 
behaviour of the eigenvalue $\gamma_1^2(\rho ) \approx -q_c \rho $ at 
$m/m_1 = \lambda_c $. 
Correspondingly, the first-channel diagonal term is of the form 
$ -\displaystyle\frac{1}{4\rho^2} - \frac{q_c}{\rho}$, where 
$q_c = \left[ \displaystyle \frac{\pi}{2}\left(1 + \frac{\pi^2}{24} - 
\frac{\omega_c^2}{2} \right) - \frac{\omega_c^3}{3\sin^2\omega_c} \right]^{-1} 
\approx 2.34253823$. 
As the energy of the $n$th level in this potential is 
$E_n = -q_c^2/(2n - 1)^2$, one can roughly estimate that at least one and not 
more than two bound states exist for $m/m_1 = \lambda_c$. 
To illustrate the above-described properties, two lowest terms 
$\gamma^2_n(\rho)/{\rho^2}$ of HREs are depicted in Fig.~\ref{figterm} for 
different values of the mass ratio. 

The asymptotic expressions for the channel potentials $V_n(\rho) = 
\displaystyle\frac{\gamma_n^2(\rho) - 1/4}{\rho^2} + P_{nn}(\rho)$ and 
the coupling terms $P_{nm}(\rho)$ and $Q_{nm}(\rho)$ are of interest for 
solution of both the eigenvalue and the scattering problem. 
The asymptotic form of $\gamma_n^2 (\rho) $ at a large hyper-radius is 
determined by the expansion of Eq.~(\ref{transeq}) for $\gamma \to i\infty$ 
and $\gamma \to 2n - 1$, which gives 
\begin{equation}
\label{as1+} 
\gamma^2_n(\rho) = \left\{
\begin{array}{l}
-\rho^2 + 2 + O(\rho^{-2}) \quad , n = 1 \ , \\ 
(2 n - 1)^2 + c_n/\rho + O(\rho^{-2}) \quad , n > 1 \ , 
\end{array} \right. 
\end{equation}
where $c_n = \displaystyle\frac{4}{\pi}\left[\frac{4n(n - 1)} 
{2n - 1} - 2 \frac{(-1)^n\cos (2n - 1)\omega}{\sin2\omega} + 
\frac{(-1)^n\sin (2n - 1)\omega}{(2n - 1)\sin^2\omega}\right]$. 
Substituting Eq.~(\ref{as1+}) in the exact 
expressions~(\ref{Qanal}),~(\ref{Panal}), and~(\ref{Pdanal}) one obtains 
a large-$\rho$ expansion for the coupling terms, 
$P_{11}(\rho) = 1/(4\rho^2) + O(\rho^{-6})$, $Q_{n1}(\rho) = O(\rho^{-5/2})$, 
$P_{n1}(\rho) = O(\rho^{-5/2})$,  $Q_{nm}(\rho) = O(\rho^{-2})$, and 
$P_{nm}(\rho) = O(\rho^{-4})$ for $n, m \ne 1$. 
The channel potentials take the asymptotic form 
\begin{equation}
\label{ef1}
V_1(\rho) = -1 + \frac{2}{\rho^2} + O(\rho^{-4}) 
\end{equation} 
and
\begin{equation}
\label{ef}
V_n(\rho) = \frac{(2n - 1/2)(2n - 3/2)}{\rho^2} + O(\rho^{-4}), 
\quad n \ge 2 \ , 
\end{equation} 
which corresponds to the long-range interaction of a dimer with the third 
particle for $n = 1$ and of three asymptotically free particles for 
$n \ge 2$. 

In addition to the above-described qualitative conclusions, a detailed 
quantitative description of the three-body properties will be given for 
the non-trivial case $a > 0$ and $m/m_1 \le \lambda_c$. 
In the following sections both the energy spectrum and the scattering 
characteristics are obtained by the numerical solution of HREs~(\ref{system1}) 
complemented by the natural zero boundary conditions $f_n(\rho) \to 0 $ as 
$\rho \to 0$ and the specified asymptotic boundary conditions as 
$\rho \to \infty$. 
All the terms of HREs are calculated by using the eigenvalue 
equation~(\ref{transeq}) and the exact expressions~(\ref{Qanal}), 
(\ref{Panal}), and~(\ref{Pdanal}) for the coupling terms, which provides 
a high accuracy of the numerical results. 

\section{Three-body bound states and near-threshold resonances}
\label{bound}

\begin{figure}[htb]
\includegraphics[width = 0.7\textwidth]{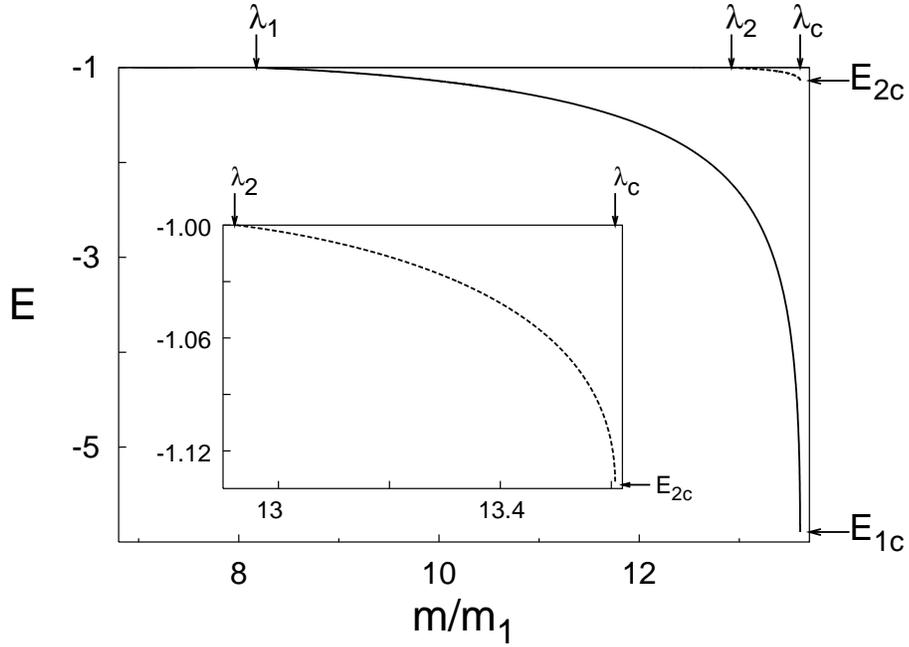}
\caption{Dependencies of the bound-state energies (in units of the two-body 
binding energy) on the mass ratio $m/m_1$. 
The arrows mark the mass ratios $\lambda_i$, for which the $i$th bound state 
emerges from the two-body threshold, the critical mass ratio $\lambda_c$, and 
the bound-state energies $E_{ic}$ for $m/m_1 = \lambda_c$. 
In the inset the excited-state energy is shown on a large scale. 
\label{E12}}
\end{figure}

\begin{figure}[htb]
\includegraphics[width = 0.44\textwidth]{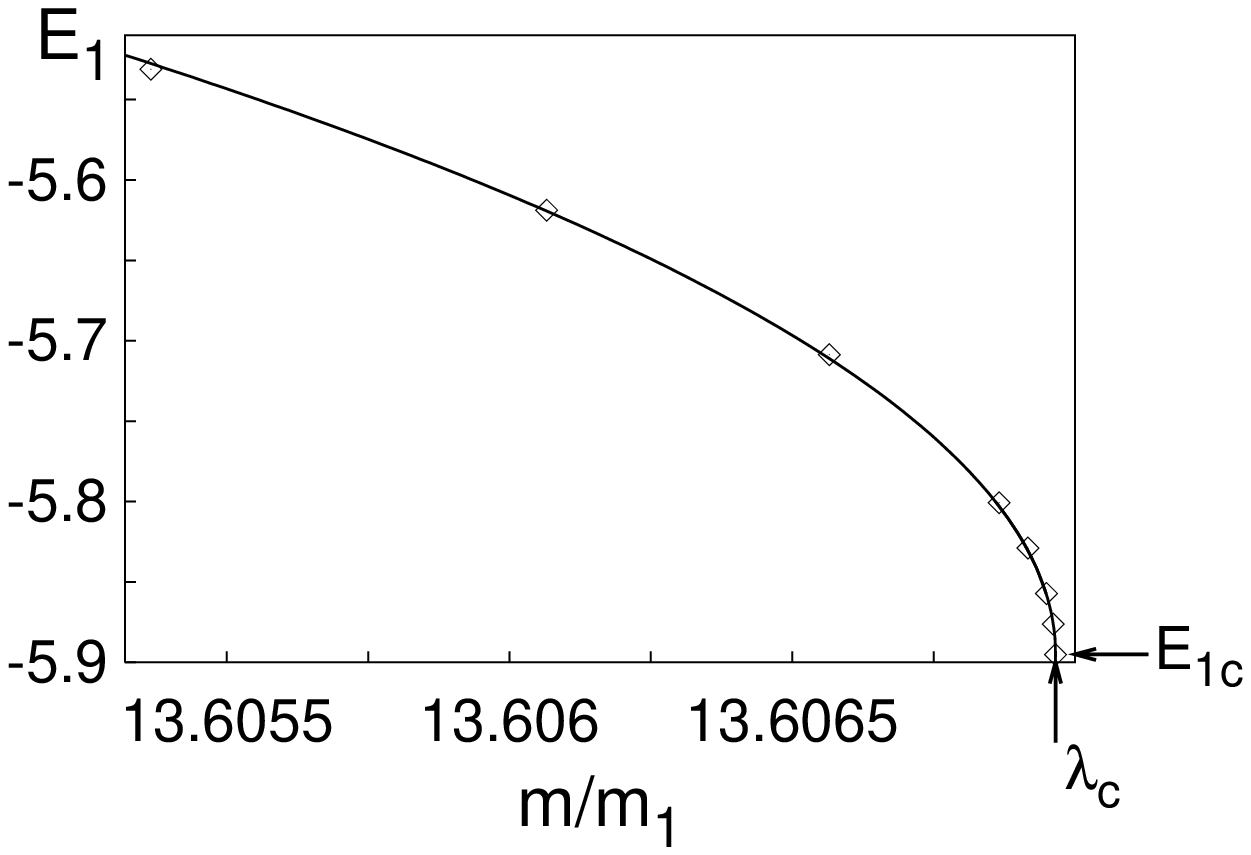}
\hspace{.9cm}
\includegraphics[width = 0.44\textwidth]{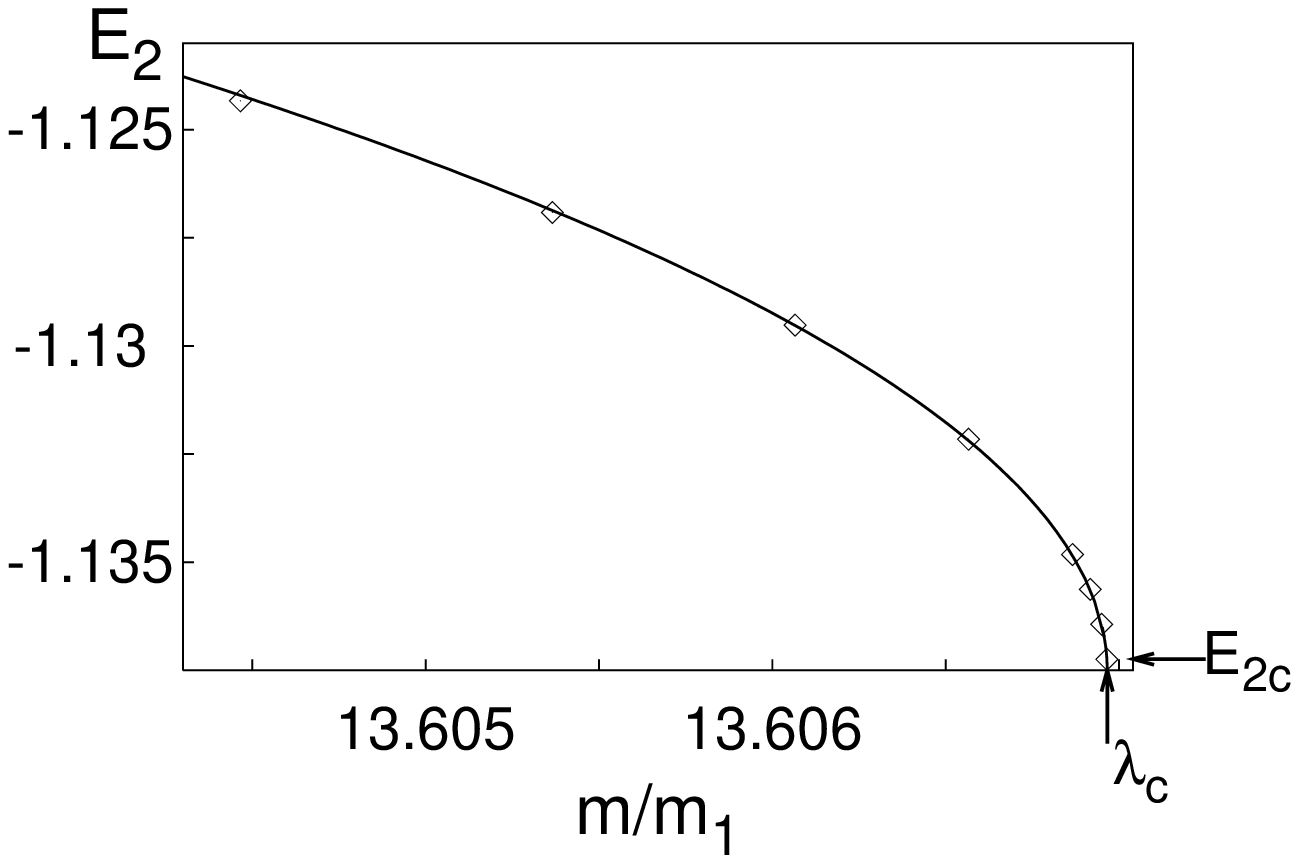}
\caption{Calculated ground-state and excited-state energies (diamonds) for 
$m/m_1 \alt \lambda_c$ fitted to the square-root dependence 
$E - E_c \propto \sqrt{\lambda_c - m/m_1}$ (lines).
\label{Ecr1}}
\end{figure}

The dependencies of the three-body binding energies on the mass ratio are 
determined by solving a system of HREs with the zero asymptotic boundary 
conditions, $f_n(\rho) \to 0 $ as $\rho \to \infty$. 
The results of the calculations are shown in Fig.~\ref{E12} and in 
Table~\ref{tab1}; it turns out that there are zero, one, and two bound states 
for $0 < m/m_1 < \lambda_1$, $\lambda_1 \le m/m_1 < \lambda_2$ and 
$\lambda_2 \le m/m_1 \le \lambda_c$, respectively. 
\begin{table}[htb]
\caption{Mass ratios $\lambda_{i}$ for which the three-body bound states arise 
and energies $E_{ic}$ of these states for $m/m_1 = \lambda_{c}$ calculated 
with $N$ HREs. } 
\label{tab1}
\begin{tabular}{ccccccccc}
$ N $ & $\lambda_{1}$ & $\lambda_{2}$ & $E_{1c}$ & $E_{2c}$ \\ 
 1 &  8.183854  & 12.929430   &  -5.89405   &  -1.13632  \\
 2 &  8.175776  & 12.921084   &  -5.89525   &  -1.13730  \\
 3 &  8.173692  & 12.918879   &  -5.89537   &  -1.13752  \\
 4 &  8.173003  & 12.918061   &  -5.89540   &  -1.13759  \\
 5 &  8.172771  & 12.917712   &  -5.89541   &  -1.13762  \\
 6 &  8.172688  & 12.917564   &  -5.89542   &  -1.13763  \\
 7 &  8.172651  & 12.917500   &     -       &     -      \\
 8 &  8.172633  & 12.917473   &     -       &     -      \\
 9 &  8.172622  & 12.917457   &  -5.89542   &  -1.13764  \\
12 &  8.172608  & 12.917436   &     -       &     -      \\
$\infty$ & 8.17260 & 12.91743 &             &            \\ 
\end{tabular}
\end{table}
The binding energies increase as the mass ratio increases to the critical 
value $\lambda_c$; in the limit $m/m_1 \to \lambda_c $ the energies tend 
to the finite values $E_{ic}$ ($i = 1, 2$) following a square-root dependence 
$E_{i} - E_{ic} \propto \sqrt{\lambda_c - m/m_1}$, which is demonstrated in 
Fig.~\ref{Ecr1}. 
Notice that this mass-ratio dependence comes from the expansion 
$\gamma_1^2(0) \propto \lambda_c - m/m_1 $ as $m/m_1 \to \lambda_c $. 

For the mass ratios $\lambda_i$, at which the three-body bound states arise, 
there are true bound states at the threshold energy $E = -1$, whose wave 
functions are square-integrable with a power fall-off at large distances. 
Thus, to calculate the precise values $\lambda_i$, a system of HREs is solved 
for $E = -1$ by using the power dependence of the first-channel function, 
$f_1(\rho) \sim \rho^{-2}$ as $\rho \to \infty$. 
The calculated $\lambda_i$ rapidly converge with increasing number of HREs 
$N$, being fairly well fitted to the power dependence $a + b/N^c$ with 
$c \approx 4$; the dependencies of $\lambda_i$ on $N$ and the fitted values in 
the limit $N \to \infty$ are presented in Table~\ref{tab1}. 
If the mass ratio slightly exceeds $\lambda_i$, the separation of the loosely 
bound state from the two-body threshold is proportional to the mass ratio 
excess, viz., $|E_i + 1| \propto m/m_1 - \lambda_i$. 
For the mass ratio just below $\lambda_i$, the relevant bound state turns to 
a narrow resonance, whose position $E^r_i$ continues a linear mass-ratio 
dependence of the bound-state energy, $E^r_i + 1 \propto \lambda_i - m/m_1$, 
whereas the width $\Gamma_i$ depends quadratically on the mass ratio excess, 
$\Gamma_i \propto (\lambda_i - m/m_1)^2$. 
The above-described threshold features are connected with the presence of 
the long-range term $2/\rho^2$ in the (2 + 1)-channel effective potential 
(illustrated in the inset of Fig.~\ref{figterm}). 

\begin{figure}[hbt]
\includegraphics[width = 0.48\textwidth]{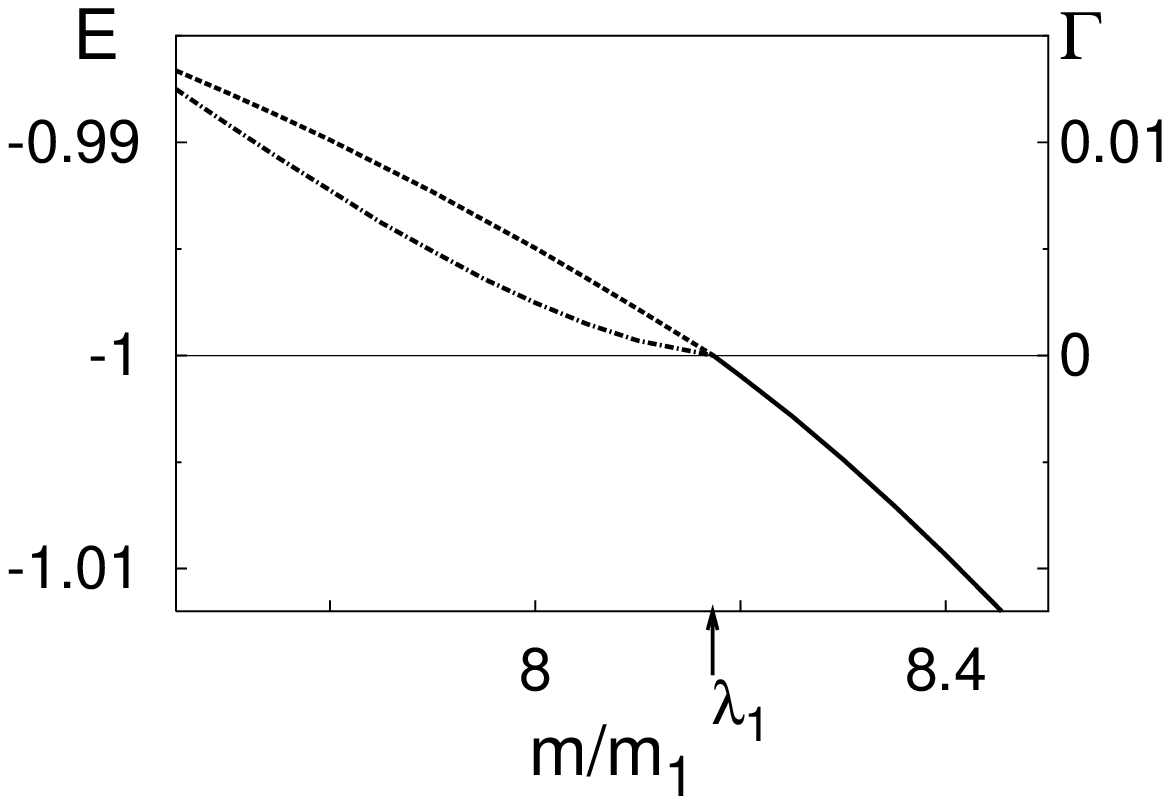}
\includegraphics[width = 0.48\textwidth]{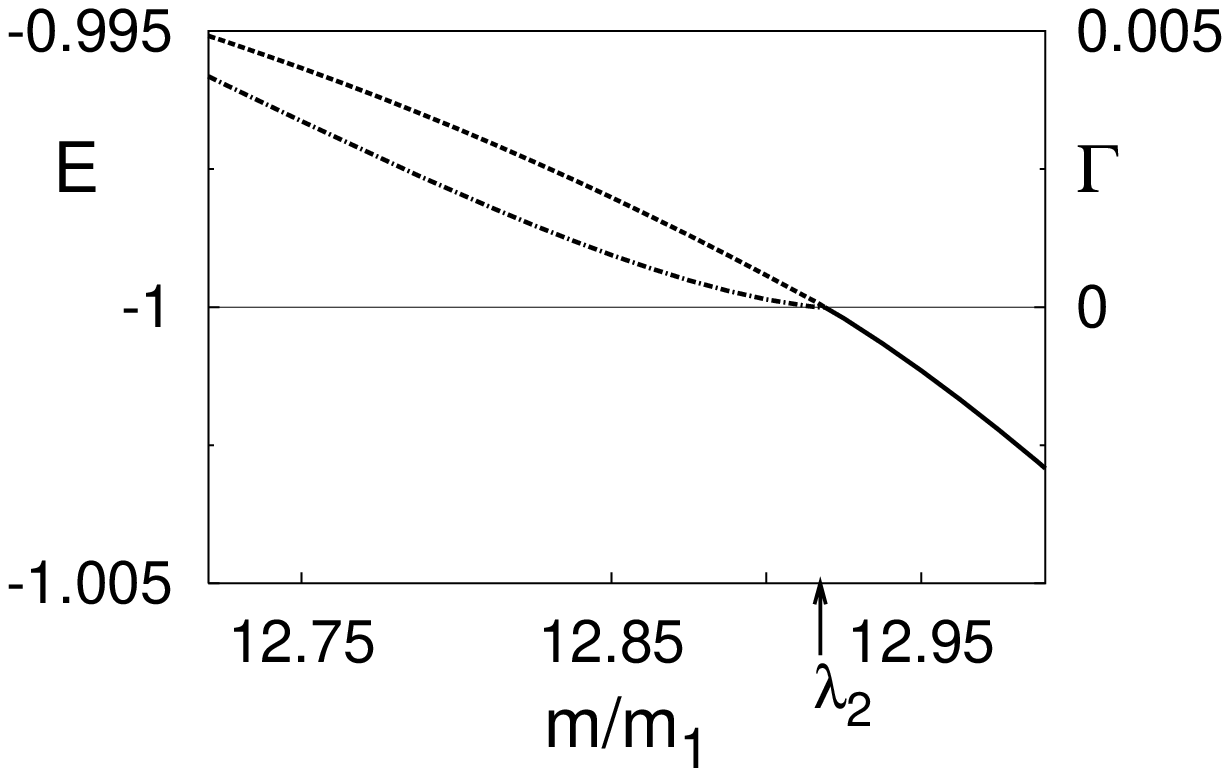}
\caption{Near-threshold mass-ratio dependencies of the bound-state energies 
$E_i$ (bold solid lines), resonance positions $E^r_i$ (thin solid lines), 
and resonance widths $\Gamma_i$ (dashed lines). 
\label{Eres12}}
\end{figure}

To calculate the positions and widths of two narrow resonances for 
$m/m_1 \alt \lambda_i$, a system of HREs is solved for $E \agt -1$. 
In view of Eq.~(\ref{ef1}), the asymptotic boundary condition for 
$\rho \to \infty $ imposed to allow for the incoming and outgoing waves in 
the first channel is taken in the form 
\begin{equation}
\label{f1e-kr}
f_1(\rho) \to \rho\left[ j_1(k\rho ) - \tan\delta(k) y_1(k\rho )\right] \ , 
\end{equation} 
where the wave number $k$ is given by $E = -1 + k^2$, $\delta(k)$ is 
the (2 + 1)-scattering phase shift, and $j_1(x)$ and $y_1(x)$ are 
the spherical Bessel functions. 
The resonance position $E_r$ and the width $\Gamma$ are determined by fitting 
$\delta(k)$ to the Wigner dependence, 
\begin{equation}
\cot [\delta(k) - \delta_{bg}] = \frac{2}{\Gamma}(E^r - E) \ , 
\label{res1}
\end{equation}
where $\delta_{bg}$ is the non-resonant phase shift. 
Near-threshold mass-ratio dependencies of the bound-state energies $E_i$ 
and the resonance parameters $E^r_i$ and $\Gamma_i$ are shown in 
Fig.~\ref{Eres12}. 

\section{Low-energy scattering near the three-body threshold }
\label{scattering}

\begin{figure}[hbt]
\includegraphics[width = 0.48\textwidth]{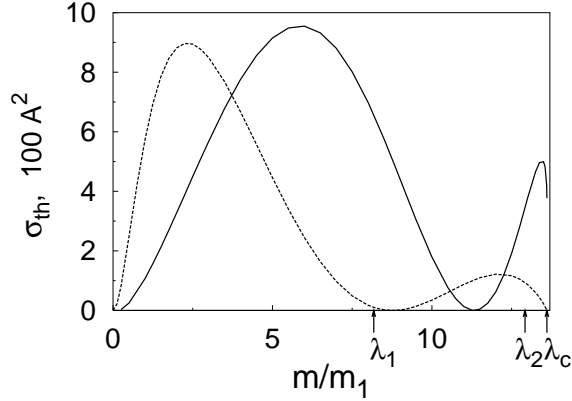}
\caption{Mass-ratio dependencies of the elastic (2 + 1)-scattering cross 
section at the three-body threshold $\sigma_{th}$ (solid line) and 
the low-energy leading-order term 
$A^2 = \displaystyle\lim_{E \to 0} |T_{21}|^2 E^{-3}$ of the three-body 
recombination rate $\alpha_r \sim A^2 E $ (dashed line). 
\label{scat}}
\end{figure}

The scattering problem at small energies near the three-body threshold 
is solved in the two-channel approximation for the mass ratio within 
the range $0 \le m/m_1 \le \lambda_c$. 
The $\mathrm{K}$-matrix is calculated by using two independent solutions 
$f^{(1)}$ and $f^{(2)}$, which satisfy, in view of Eq.~(\ref{ef1}) 
and~(\ref{ef}), the following asymptotic boundary conditions 
\begin{eqnarray}
\left[ f^{(1)},f^{(2)} \right] = \sqrt{\rho} 
\left[ 
\begin{pmatrix}
Y_{3/2}(k\rho) & 0 \\
0 & Y_{3}(\sqrt{k^2 - 1}\rho)
\end{pmatrix} + \mathrm{K} 
\begin{pmatrix}
J_{3/2}(k\rho) & 0 \\
0 & J_{3}(\sqrt{k^2 - 1}\rho)
\end{pmatrix}
\right] 
\end{eqnarray} 
as $\rho \to \infty$. 
The elastic (2 + 1)-scattering phase shift is defined by 
$\cot\delta(k) = - K_{11}(k)$ and the inelastic scattering amplitude is 
determined by the non-diagonal element of the $\mathrm{T}$-matrix given by 
$\mathrm{T} = 2(1 - i\mathrm{K})^{-1}$. 
The elastic-scattering phase shift at the three-body threshold 
$\delta_{th} \equiv \delta(1)$ is a smooth increasing function of $m/m_1$, 
which takes the value between $3\pi/2$ and $2\pi$ at $m/m_1 = \lambda_c$. 
Correspondingly, the elastic-scattering cross section at the three-body 
threshold is determined in the dimensional units as $\sigma_{th} = 
12\pi a^2 \displaystyle\frac{(1 + 2 m/m_1)}{(1 + m/m_1)^2} \sin^2\delta_{th}$. 
As shown in Fig.~\ref{scat}, the mass-ratio dependence of $\sigma_{th}$ is 
a two-hump structure with two maximums located near those values $m/m_1$ at 
which $\delta_{th}$ passes through $\pi/2$ and $3\pi/2$. 

The low-energy dependence of the three-body recombination rate $\alpha_r$ is 
determined by the squared non-diagonal $\mathrm{T}$-matrix element as 
$\alpha_r \sim |T_{21}(E)|^2 E^{-2}$. 
For small $E \to 0$, the numerical calculations corroborate the dependence 
$T_{21}(E) = E^{3/2} (A + B E)$, which agrees with the linear low-energy 
behaviour $\alpha_r \sim A^2 E $ described in Refs.~\cite{Petrov03,Suno03}. 
The mass-ratio dependence of the leading-order term $A^2$, which determines 
the three-body recombination rate at low energy, is shown in Fig.~\ref{scat}. 
A two-hump structure of $A^2(m/m_1)$ with two maxima and three zeros within 
the interval $0 \le m/m_1 \le \lambda_c$ is in agreement with the result of 
Ref.~\cite{Petrov03}. 

A similar structure of both dependencies $\sigma_{th}(m/m_1)$ and $A^2(m/m_1)$ 
originates from the interference of the incoming and outgoing waves in 
the 2 + 1 channel, being closely connected with the potential-well deepening 
as $m/m_1$ increases. 
Qualitatively, the (2 + 1)-channel function $f_1(\rho)$ acquires additional 
oscillations as the potential-well depth increases, which leads to 
an oscillating mass-ratio dependence of both elastic and inelastic scattering 
amplitudes. 
On the other hand, arising of the three-body bound states with increasing 
$m/m_1$ is connected with occurrence of oscillations of $f_1(\rho)$ within 
the potential-well range. 
In this respect, one can mention an analogy with Levinson's theorem, which 
links the number of the bound states to the threshold-energy phase shift. 

\section{Discussion}
\label{Discussion}

The universal low-energy description for two identical fermions interacting 
with the third different particle in the states of the total angular momentum 
$L = 1$ is given within the framework of the approach based on the solution of 
hyper-radial equations, whose terms are derived in the analytical form. 
It is found that there are no three-body bound-states for the negative 
scattering length and $m/m_1 \le \lambda_c$, whereas for the positive 
scattering length there are exactly zero, one and two bound states for 
$m/m_1 < \lambda_1$, $\lambda_1 \le m/m_1 < \lambda_2$ and 
$\lambda_2 \le m/m_1 \le \lambda_c$, respectively. 
For $m/m_1$ just below $\lambda_1$ or $\lambda_2$, the bound states disappear 
and turn to narrow resonances, whose positions and widths are calculated. 

The above-described universal picture should be observed in the limit 
$|a| \to \infty$, i.~e., if the potential is tuned to produce the loosely 
bound two-body state. 
In this limit, one expects to observe simultaneously the loosely bound 
two-body and three-body states, whose binding energies scale as $a^{-2}$ and 
their ratio depends on $m/m_1$. 
Similar threshold behaviour of the binding energies was discussed 
in~\cite{Cabral79,Blume05,Kartavtsev06} for three two-dimensional bosons. 

Both the elastic (2 + 1)-scattering cross sections and the three-body 
recombination rate near the three-body threshold manifest a two-hump structure 
of their mass-ratio dependencies for $m/m_1 \le \lambda_c$. 
The structure of both isotopic dependencies stems from the interference of 
the incoming and outgoing waves due to deepening of the effective potential in 
the 2 + 1 channel; in this respect, the interference is connected with arising 
of two three-body bound states with increasing $m/m_1$. 

As the present paper describes the universal three-body properties in 
the idealized limit of the zero interaction range, it is of interest to 
discuss briefly the effect of the finite, though small enough interaction 
radius $r_0 \ll a$. 
For the mass ratio below the critical value $\lambda_c$, the binding energies 
depend smoothly on the interaction radius $r_0$ and on the interaction in 
the vicinity of the triple-collision point provided $r_0 \ll a$, whereas for 
$m/m_1 > \lambda_c$ the infinite energy spectrum is extremely sensitive to 
these parameters. 
Furthermore, an abrupt transition from two to an infinite number of bound 
states at $m/m_1 = \lambda_c$ will be smeared off if either the interaction 
range is not zero or the three-body force is present. 
One can roughly estimate that the number of three-body bound states 
$N_b = 2 $ for the mass ratio within the range $m/m_1 - \lambda_c \alt r_0/a$ 
and increases as $N_b \propto \sqrt{m/m_1 - \lambda_c} \ln\frac{a}{r_0}$ with 
increasing $m/m_1$. 

Finally, it is worth noting that the p-wave trimer molecule containing two 
heavy fermions and the light third particle could be observed in 
the ultra-cold mixtures of $^{87}\mathrm{Sr}$ with lithium isotopes. 
For the mixtures of $^{87}\mathrm{Sr}$ with $^7\mathrm{Li}$, the mass ratio 
$m/m_1 \approx 12.4 > \lambda_1$, which entails existence of the trimer 
bosonic molecule $^7\mathrm{Li} \, ^{87}\mathrm{Sr}_2$, whose binding energy 
is about $0.793$ times the binding energy of the dimer molecule 
$^7\mathrm{Li} \, ^{87}\mathrm{Sr}$. 
For the mixtures of $^{87}\mathrm{Sr}$ with $^6\mathrm{Li}$, the mass ratio 
$m/m_1 = 14.5$ slightly exceeds the critical value $ \lambda_c$, which entails 
existence of the trimer fermionic molecule 
$^6\mathrm{Li} \, ^{87}\mathrm{Sr}_2$ in two states, whose binding energies 
are slightly above $4.895$ and $0.138$ times the binding energy of the dimer 
molecule $^6\mathrm{Li} \, ^{87}\mathrm{Sr}$. 

{\em Acknowledgement.}
Support by the grant of "Econatsbank" and the Votruba-Blokhintsev grant is 
gratefully acknowledged. 

\appendix
\section{Analytical expressions for the coupling terms}

Although the direct calculation of the coupling terms $Q_{nm}(R)$ and 
$P_{nm}(R)$ via the definition~(\ref{QPnm0}) is quite involved, one can 
circumvent this problem and obtain the analytical expressions for 
$Q_{nm}(\rho)$ and $P_{nm}(\rho)$ via $\gamma_n^2(\rho)$ and their derivatives 
by using the explicit dependence on $\rho$ in the boundary 
condition~(\ref{bch}). 
Similar analytical expressions were derived for a number of problems based on 
the BCM; more details are given in~\cite{Kartavtsev06}. 

Hereafter one concisely writes the eigenvalue 
problem~(\ref{eqonhypershere}),~(\ref{bch}) as 
\begin{eqnarray}
\label{eigv}
\left( \tilde\Delta + \gamma_n^2 \right) \Phi_n = 0 \ , \\ 
\label{eigvbc}
\lim_{\alpha \rightarrow 0} \left( \frac{\partial}{\partial\alpha} + 
\rho \right) \sin 2\alpha \Phi_n = 0 \ . 
\end{eqnarray} 
The derivative of the normalized eigenfunction $\Phi_n $ with respect to 
$\rho$ satisfies the inhomogeneous equation 
\begin{equation}
\label{eigvder}
\left( \tilde\Delta + \gamma_n^2 \right) \frac{\partial\Phi_n}{\partial\rho} + 
\frac{d \gamma_n^2}{d \rho} \Phi_n = 0 
\end{equation} 
and the boundary condition 
\begin{equation}
\label{eigvderbc}
\lim_{\alpha\rightarrow 0} \left[ \left( \frac{\partial}{\partial\alpha} + 
\rho \right) \sin 2\alpha \frac{\partial\Phi_n}{\partial\rho} + 
\sin 2\alpha \Phi_n \right] = 0 \ . 
\end{equation}

By projecting Eq.~(\ref{eigvder}) onto $\Phi_m $ and using 
the representation~(\ref{faddeev}) one obtains the relation, 
\begin{equation}
\label{scalprod}
(\gamma_n^2 - \gamma_m^2)Q_{mn} + \delta_{nm}\frac{d \gamma_n^2}{d \rho} + 
\varphi_n(0, \rho)\varphi_m(0, \rho) = 0 \ , 
\end{equation}
where the integrals over the hypersphere are expressed via the integrals over 
the hyper-surfaces surrounding two singularities of the functions $\Phi_n$, 
viz., one at $\alpha = 0 $ and the other given by the permutational symmetry. 
Here the integration volume is arbitrarily chosen to provide the unit 
coefficient for the last term in~(\ref{scalprod}) and it is taken into account 
that equal contributions come from two surface integrals around both 
singularities. 
The diagonal part of Eq.~(\ref{scalprod}) gives the basic relation 
\begin{equation}
\label{varphi0}
\varphi_n^2(0, \rho) = -\displaystyle\frac{d \gamma_n^2}{d \rho} \ , 
\end{equation} 
which allows one to derive the desired expressions via the derivative of 
the eigenvalues $\gamma_n^2(\rho)$. 
Substituting~(\ref{varphi0}) in the non-diagonal part of~(\ref{scalprod}), one 
finds 
\begin{equation}
\label{Qanal}
Q_{nm} = \left( \gamma_n^2 - \gamma_m^2 \right)^{-1} 
\sqrt{\frac{d \gamma_n^2}{d \rho} \frac{d \gamma_m^2}{d \rho}}  \ . 
\end{equation}
In a similar way, the projection of Eq.~(\ref{eigvder}) onto $\displaystyle 
\frac{\partial\Phi_m}{\partial\rho}$ for $n \ne m$ leads to the relation 
\begin{equation}
\label{scalprod1}
\frac{d (\gamma_n^2 + \gamma_m^2)}{d \rho} Q_{mn} = 
(\gamma_n^2 - \gamma_m^2) P_{mn} + 
\varphi_n(0, \rho)\frac{d \varphi_m(0, \rho)}{d \rho} - 
\varphi_m(0, \rho)\frac{d \varphi_n(0, \rho)}{d \rho} \ , 
\end{equation}
which is finally transformed to the expression for the non-diagonal coupling 
terms 
\begin{equation}
\label{Panal}
P_{nm} = Q_{nm} \left[ \left( \gamma_m^2 - \gamma_n^2 \right)^{-1} 
\frac{d }{d \rho} \left(\gamma_n^2 + \gamma_m^2 \right) + 
\frac{1}{2}\frac{d^2 \gamma_n^2}{d \rho^2} 
\left(\frac{d \gamma_n^2}{d \rho}\right)^{-1} - 
\frac{1}{2}\frac{d^2 \gamma_m^2}{d \rho^2} 
\left(\frac{d \gamma_m^2}{d \rho}\right)^{-1} \right] \ , 
\end{equation}
where one uses Eq.~(\ref{varphi0}) and its derivative 
$\displaystyle \frac{d^2 \gamma_n^2}{d \rho^2} = - 
2 \varphi_n(0, \rho) \frac{d \varphi_n(0, \rho)}{d \rho}$. 

At last, the second derivative of the eigenfunction $\Phi_n$ with respect to 
$\rho$ satisfies the equation 
\begin{equation}
\label{eigvder2}
\left(\tilde\Delta + \gamma_n^2 \right)\frac{\partial^2\Phi_n}{\partial\rho^2} 
+ 2 \frac{d \gamma_n^2}{d \rho} \frac{\partial\Phi_n}{\partial\rho} + 
\frac{d^2 \gamma_n^2}{d \rho^2} \Phi_n = 0 
\end{equation}
and the boundary condition 
\begin{equation}
\label{eigvder2bc}
\lim_{\alpha\rightarrow 0} \left[ \left( \frac{\partial}{\partial\alpha} + 
\rho \right) \sin 2\alpha \frac{\partial^2\Phi_n}{\partial\rho^2} + 
2 \sin 2\alpha \frac{\partial \Phi_n}{\partial\rho} \right] = 0 \ . 
\end{equation}
By projecting Eq.~(\ref{eigvder2}) onto $\Phi_n $ and using the identity 
$P_{nn} = -\left\langle \Phi_n \Bigg| \displaystyle 
\frac{\partial^2 \Phi_n}{\partial\rho^2} \right\rangle $, one finds that 
\begin{equation}
\label{scalprod2}
3 \frac{d \gamma_n^2}{d \rho} P_{nn} = 
\varphi_n(0, \rho)\frac{d^2 \varphi_n(0, \rho)}{d \rho^2} - 
2 \left[\frac{d \varphi_n(0, \rho)}{d \rho}\right]^2 \ . 
\end{equation}
The derivatives of $\varphi_n(0, \rho)$ are expressed via the derivatives of 
$\gamma_n^2(\rho)$ by using Eq.~(\ref{varphi0}), which allows one to cast 
the diagonal coupling term in the form 
\begin{equation}
\label{Pdanal}
P_{nn} = -\frac{1}{6}\frac{d^3 \gamma_n^2}{d \rho^3} 
\left(\frac{d \gamma_n^2}{d \rho}\right)^{-1} + 
\frac{1}{4}\left(\frac{d^2 \gamma_n^2}{d \rho^2}\right)^2 
\left(\frac{d \gamma_n^2}{d \rho}\right)^{-2} \ .
\end{equation}

\bibliography{fermions}

\end{document}